# Towards Realisation of Heterogeneous Earth-Observation Sensor Database Framework for the Sensor Observation Service based on PostGIS

Maduako N. Ikechukwu & Francis I. Okeke

Department of Geoinformatics and Surveying,

University of Nigeria, Enugu Campus

Enugu, Nigeria

**Abstract**

*Environmental monitoring and management systems in most cases deal with models and spatial analytics that involve the integration of in-situ and remote Geosensor observations. In-situ sensor observations and those gathered by remote sensors are usually provided by different databases and services in real-time dynamic services such as the Geo-Web Services. Thus, data have to be pulled from different databases and transferred over the network before they are fused and processed on the service middleware. This process is very massive and unnecessary communication-work load on the service middleware. Massive work load in large raster downloads from flat-file raster data sources each time a request is made and huge integration and geo-processing work load on the service middleware which could actually be better leveraged at the database This paper therefore proposes the realization of heterogeneous sensor database framework based on PostGIS for integration, geo-processing and spatial analysis of remote and in-situ sensor observations at the database level. Also discussed in this paper is how the framework can be integrated in the Sensor Observation Service (SOS) to reduce communication and massive workload on the Geospatial Web Services and as well make query request from the user end a lot more flexible.*

## 1. Introduction

Geo-sensors gathering data to the geospatial sensor web can be classified into remote sensors and in-situ sensors. Remote sensors include satellite sensors, Unmanned Aerial Vehicles (UAV), Light Detection and Ranging (LIDAR), Aerial Digital Sensors (ADS) and so on, measuring environmental phenomena remotely. These sensors acquire data in raster format at larger scales and extent. In-situ sensors are spatially distributed sensors over a region used to monitor and observe environmental conditions such as temperature, sound intensity, pressure, pollution, vibration, motion etc. These sensors are measuring phenomena in their direct environment and could be said to acquire data in vector data format.

Most environmental monitoring and management systems combine these disparate datasets from heterogeneous sensors for environmental modeling and analysis. For example in monitoring of crop Actual Evapotranspiration at some locations in most cases involves aggregation of remote and in-situ sensor observations (Gregory Cutrell, 2009). Remote and in-situ sensor data aggregation for real-time calculation of daily crop Gross Primary Productivity GPP such as implemented in a dynamic web mapping service for vegetation productivity (Kooistra, Bergsma, Chuma, & de Bruin, 2009)and in the marine information system (Hamre, 1993)are good examples too.

Meanwhile the process of fusing and processing these sensor data on the web service currently involves massive data retrieval from different sensor databases, geo-processing and spatial analytics on service middleware. For web services, this entails massive work and communication load over the network and on the service. A sensor database management framework combining remote and in-situ observations would be of great impact to environmental monitoring and management systems. Having these disparate sensor data on one database schema can be leveraged in the geospatial web services to reduce excessive work load and data transfer through the network. Most of the data fusion, aggregations and processing done by web services can be carried out at the database backend and the results delivered to the client through the appropriate web services.

Figure 1 below, is a diagrammatical illustration of this approach. In-situ and remote sensor data are collected at the heterogeneous sensor database, integration and processing carried out at the database level within the SOS and geo-scientific query results are delivered to the clients through the service.

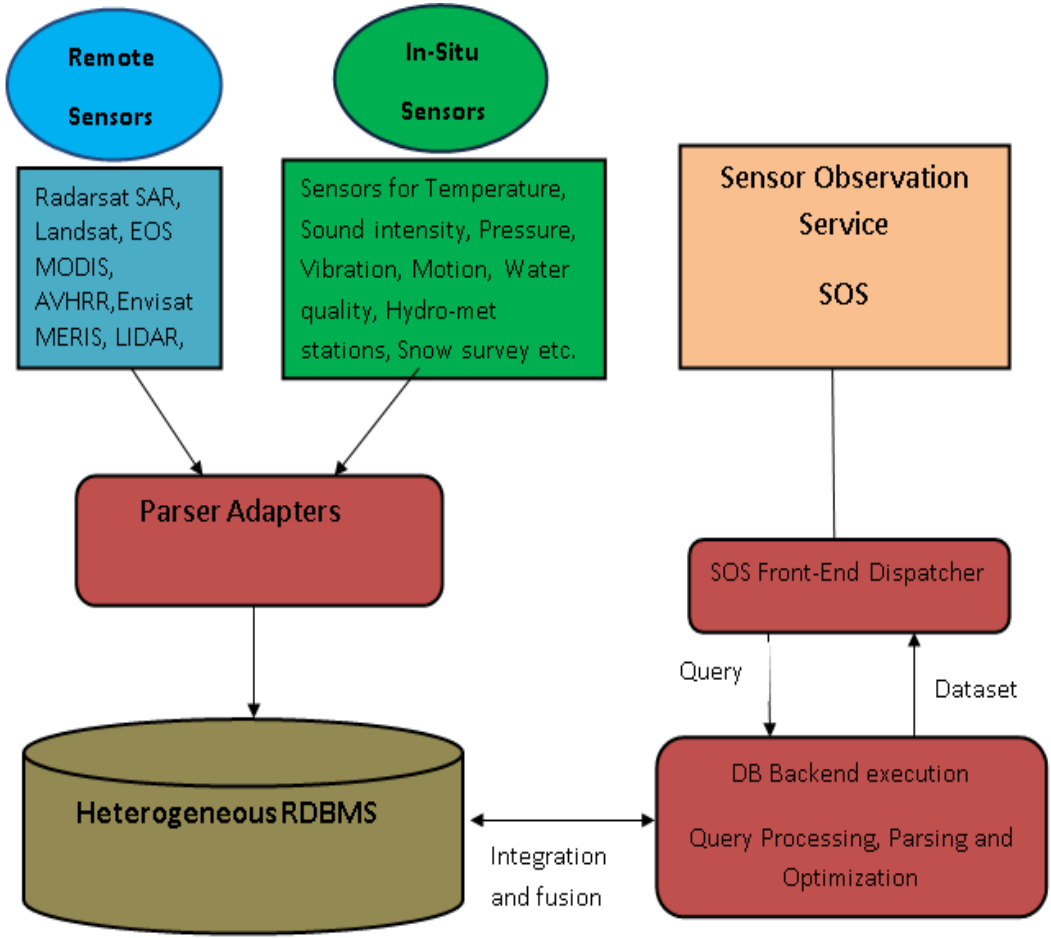

Figure 1 : The Conceptual Diagram

## 2. *Requirement Analysis of this Aproach*

Firstly, analysis of the fundamental conceptual and practical requirements for the proposed heterogeneous sensor database framework for a seamless integration of remote and in-situ sensor observations at the database level of the SOS was carried out. The analysis is done taking into consideration the varying properties and the underlying structure and format of these two different sensor datasets (raster and vector). The database model for this purpose can be design as a spatial database model based on the Open Geospatial Consortium (OGC) standards. Sensor observations can be approached in perspective of coverages. This is to say we can treat in-situ sensor observations as time series vector coverage and remote sensor observations as also time dependent raster coverage.

Coverages have some fundamental properties, exploring some of these properties and how vector and raster coverages inherit these properties, we can conceptually map out an intersection that will underline the seamless integration of vector and raster (in-situ and remote sensors) coverages in a heterogeneous sensor database schema, see figure 2.

> According to ISO 19123: 2005 "*a coverage domain consists of a collection of direct positions in a coordinate space that may be defined in terms of up to three spatial dimensions as well as a temporal dimension"* (ISO, 2009).

A coverage is created as soon as a way to query for a certain value given a location is created.

Coverages can be categorised into two, continuous and discrete coverages. Continuous coverage returns a value of the phenomenon at every possible location within the domain. Discrete coverages can be derived from the discretisation of a continuous surface. A discrete coverage consists of different domain and range sets. The domain set consists of either spatial or temporal geometry objects, finite in number. The range set is comprised of a finite number of attribute values each of which is associated to every direct position within any single spatio-temporal object in the domain. That is to say, the range values are constant on each spatio-temporal object in the domain. *"Coverages are like mathematical functions, they can calculate, lookup, intersect, interpolate, and return one or more values given a location and/or time. They can be defined everywhere for all values or only in certain places ''* (PostGISWiki)*.*

Raster and vector coverages are both types of discrete coverage. They differ only in how they store and manage their collection of data.  As coverages, they allow for basic query functions such as select, find, list etc. to be carried out on them.

Vector coverages handled as tables are the most common type of coverage implemented in most of the spatial database management systems. Individual data item are stored on each row in the table. The columns of the table ensure that collection is self-consistent. Texts are placed in text columns, numbers in numeric columns, and geometries in geometry columns and so on. The basic requirement a table must have for potential supply of information to a coverage is to have at least one geometry column and one additional column for a value or an attribute.

Raster coverages are handled as an array of multidimensional discrete data as discussed in (Baumann & Peter, 1994)In PostgreSQL/PostGIS 2.0 (Racine, 2011)precisely they are stored as regularly gridded data with the geometry of the domain as points and the range could be one or more numeric values (for example number of bands). Text values and timestamps may not be possible.

Hence, in-situ observations (vector data) can be stored in tables with rows and columns in a relational manner, having one-to-one or one-to-many relationship. On the other hand remote sensor observations (raster) cannot reasonably be stored in tables but as gridded multidimensional array of data (array of points). That is to say we only have to leverage the concept of coverage to integrate the two tables in the database. The possible common column for the two datasets (tables) is the geometry column.

Therefore the fundamental requirement from this analysis that could enable us to integrate remote and in-situ sensor observations in a common database could be outlined as:

- storage of in-situ observations as vector point coverage and
- storage remote sensor observations as raster point (pixel) coverage.

Figure 2 is the UML (Unified Modeling Language) model of the concept and management of coverages describing features, relationships, functions and how they present in the database. The insight to this UML model was extracted from the coverage concept model discussed in

PostGIS Wiki - (5).

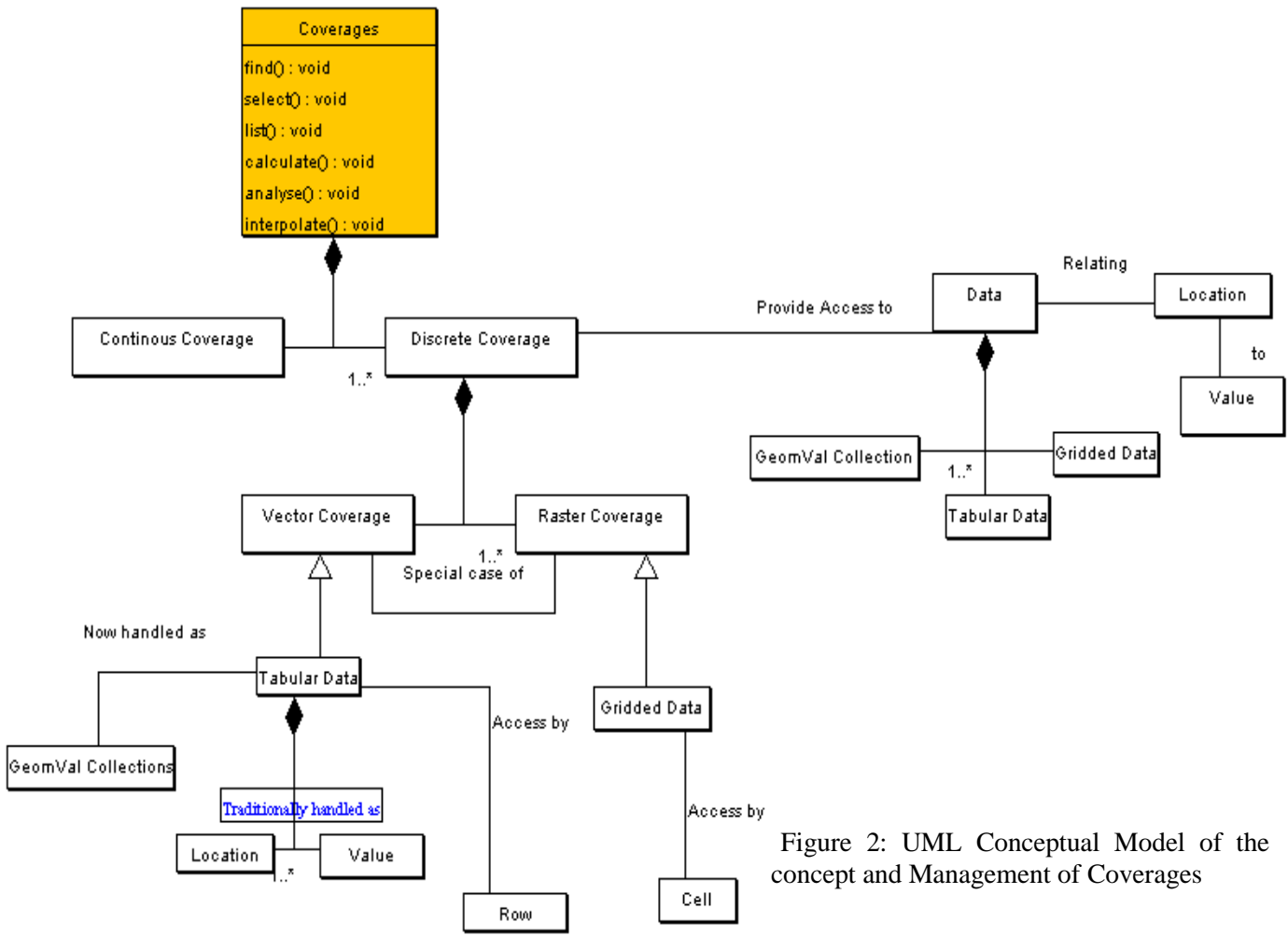

Figure 2: UML Conceptual Model of the concept and Management of Coverages

Leveraging these functions and operations that can be carried out on coverages, the database can offer fundamental operations and functions such as intersection, buffering, overlay, interpolation etc for geo-scientific analysis and processing involving in-situ and remote sensor observations (vector and raster coverages).

With these operations, we can easily run queries for example that can lift a point on the vector coverage, intersect it with the geometrically corresponding point or cell on the raster coverage on the database and return a value. The goal is for us to be able to do relation and overlay operations on the different coverages irrespective of how the coverages are stored. Therefore we need a database management system that can provide these supports for this purpose.

*2.1 Database Management System (DBMS) Support Analysis*

Effective storage and retrieval of vector data has been well developed and implemented in most of the spatial databases such as Postgresql/PostGIS, Oracle Spatial, MySQL, Microsoft SQL Server 2008, SpatiaLite, Informix, etc. On the other hand, Oracle Spatial and Postgresql/PostGIS DBMS are currently the only DBMS that have substantial support for raster data management. Meanwhile Oracle Spatial supports raster data storage with less support for raster data analysis in the database. However PostgreSQL/PostGIS2.0 has relatively good raster support, functions and operations that we can leverage for the feasibility of our research goal. In addition PostgreSQL/PostGIS2.0 can be configured with python GDAL-bonded to leverage more functionality (PostGISWiki).

PostgreSQL/PostGIS2.0 capability to carry out seamless vector and raster data integration makes it favourable in this type of our work than Oracle Spatial. PostgreSQL/PostGIS2.0 can handle pixel-level raster analysis unlike Oracle Spatial whose content search is based on Minimum Bounding Rectangle (MBR). PostgreSQL/PostGIS2.0 uses Geospatial Data Abstraction Libraries (GDAL) to handle multi-format image input and output and when working with out-db-raster, this is a powerful functionality (PostGISWiki).

PostgreSQL/PostGIS2.0 supports GiST spatial indexing, GiST stands for "Generalized Search Tree" and is a generic form of indexing. GiST is used to speed up searches on all kinds of irregular data structures (integer arrays, spectral data, etc) which are not amenable to normal B-Tree indexing (postgis.refractions.net)

In PostgreSQL/PostGIS2.0, raster coverage can be created by having a geometry column called raster and attribute columns containing the attributes to the raster (e.g. band number, timestamp and so on). The fundamental database or storage support needed on the raster coverage for efficient seamless integration and analysis with vector coverages such as tiled raster storage, georeferencing, multiband/multi-resolution support and so on are provided by PostgreSQL/PostGIS2.0 (Racine, 2011). Structured Query Language (SQL) raster functions and operators for raster manipulations and analysis are substantially supported in PostgreSQL/PostGIS2.0, more functions are being developed.

3. *Conceptual Design and Modelling of a Heterogeneous Database Schema*

Based on those fundamental requirement analysis, the conceptual model of a heterogenous sensor database schema, integrating remote and in-situ sensor observations was developed. The UML model in figure 3 shows the high level abstraction model of the fundamental classes (tables) that are needed in a sensor database, their attributes and important operations that can be carried out on them. It shows the relationships and the logic between the classes which enable integration between the classes. The Entity Relationship (ER) diagram in figure 4 describes the logical design for physical implementation of the entities, the fields in each class and the relationships between entities. Also in this section we developed the conceptual model of how the database model can be integrated with other web services seamlessly, introducing the concept of the Web Query Service, WQS.

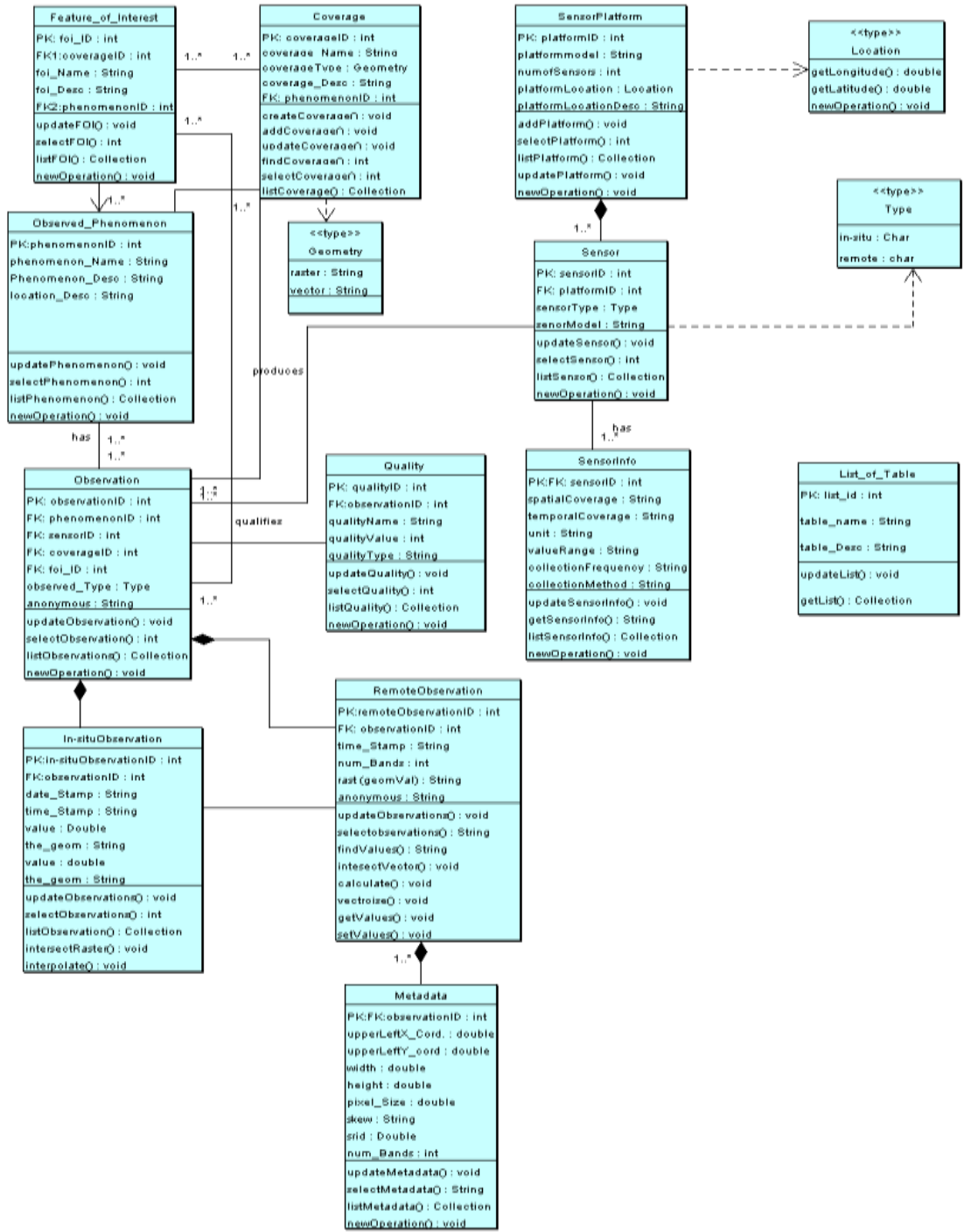

Figure 3: UML Conceptual Schema Model of the Proposed Heterogeneous Sensor Database

*3.1 The ER-diagram and logical design of the database model*

Figure 4 is the Entity Relationship diagram and logical design of the proposed heterogenous sensor database. The diagram shows the relationship logics between the tables for an effective physical implementation in PostGIS, leveraging the primary and foreign keys for seamless integration between the class. The relationship and integration of the In-situObservation and RemoteObservation tables are executed on the fly, leveraging their geometry columns and the coverage concept .

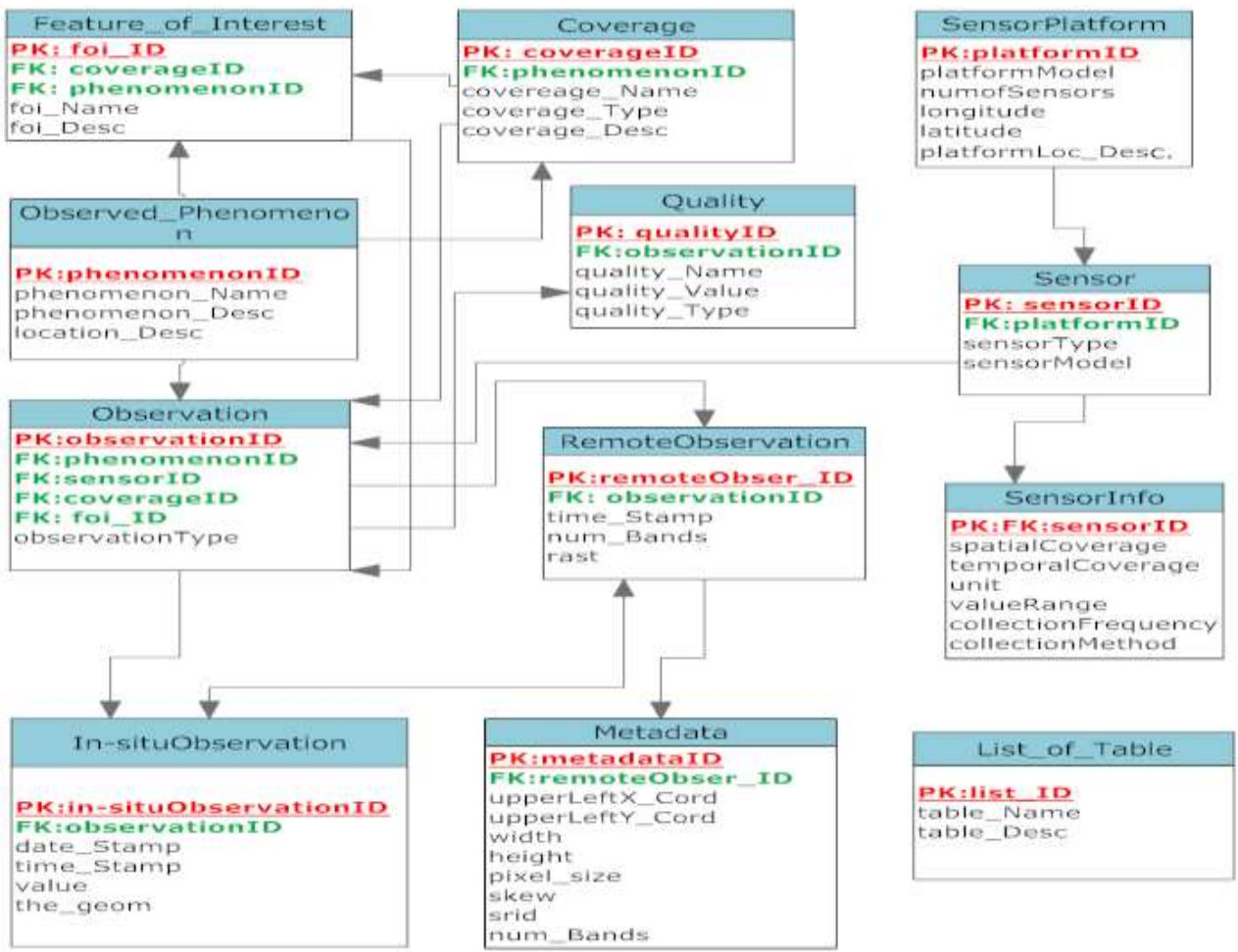

Figure 4: ER-Diagram and Logical Design of the Heterogeneous Database Mode

## 4. Integrating the Heterogeneous Sensor Database with the OGC Web Services

We propose an SQL-based Web Query Service, WQS that delivers SQL queries from the user end to the database in the web service . This service can be intergated and accessed from within the user's web or desktop application. This service provides the cleint the flexibility and ability to construct queries of extensive complexity which is delivered to the database for processing. In this case, aggregations, processing and analysis of remote and in-situ observations are carried out at the database backend. The result of the query can be delivered in different formats such as ASCII, GML, KML, TIFF, JPEG etc. in compliance with the OGC web mapping services, the WFS, WMS and WCS. The user specifies the formats of delivery on the query by using the PostGIS "ST_As*" function. ASCII or text results are delivered to the client directly from the database through the WQS. If the request result is to be delivered as a raster coverage, then the query result is a raster or a rasterised vector and will be delivered to the client through the Web Coverage Service, WCS protocol. Similar process goes for a vector or vectorised query result which is delivered through the Web Feature service WFS protocol. The request result can be delivered as a JPEG or PNG image format to the user through the Web Map Service WMS protocol as described in figure 6.

*4.1 The concept of the Web Query Service WQS*

The Web Query Service, WQS is the proposed SQL query service that serves query from the client's web or desktop application to the heterogeneous sensor database. The Web Query Service delivers SQL queries from the client application through the network to the sensor database. It makes it easier to build and execute queries on a remote sensor database from any client application.

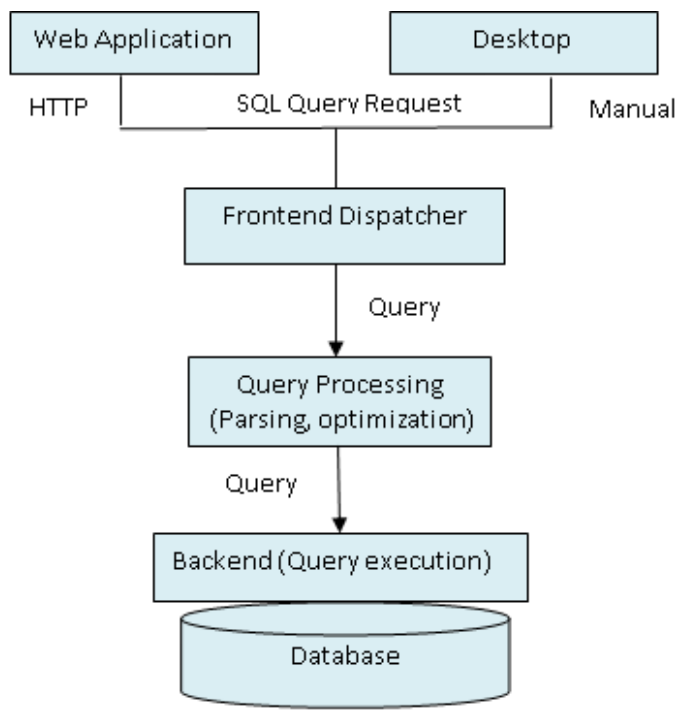

Figure 5: Conceptual Model of the proposed Web Query Service WQS

In Figure 5 the SQL query is delivered from the frontend dispatcher of client web or desktop application to the query processing and optimization module for optimization and parsing to the backend for query execution. From a web application, the SQL query request is dispatched via the HTTP. From within a desktop application, a connection to the database would have to be established manually before queries are sent to the database for execution.

**4.2 Conceptual Architecture for the Integration of the Heterogenous Sensor Database with OGC Web Services.**

Figure 6 describes the conceptual achitecture of our proposed integration of the heterogenous sensor database as part of the Sensor Observation Service with the proposed Web Query Service WQS and other Web Services to deliver effective results to the end user.

The user on the client end, web or desktop application delivers SQL queries of any complexity through the WQS to the database. The result of the query is delivered back to the user through the relevant services depending on the format the result is requested. The ST_As * PostGIS function is used in the query to specify the format of delivery. When the user specifies for example ST_As GeoTIFF, the raster coverage query result is wrapped in XML and delivered to the client through the WCS protocol. The same process goes for query results specified in ST_As JPEG, PNG and KML or GML which are dilivered through the WMS and WFS respectively to the client. If no delivery format is specified in the query, the result is returned back to the client via the WQS by defualt in ASCII format. OGC web service operations such as GetCapabilities, DescribeSensor, DescribePlatform, GetObservation, DescribeCoverage or GetRaterMetadata, GetCoverage, ProcessCoverage etc. are carried out through this Web Query Service WQS by SQL queries.

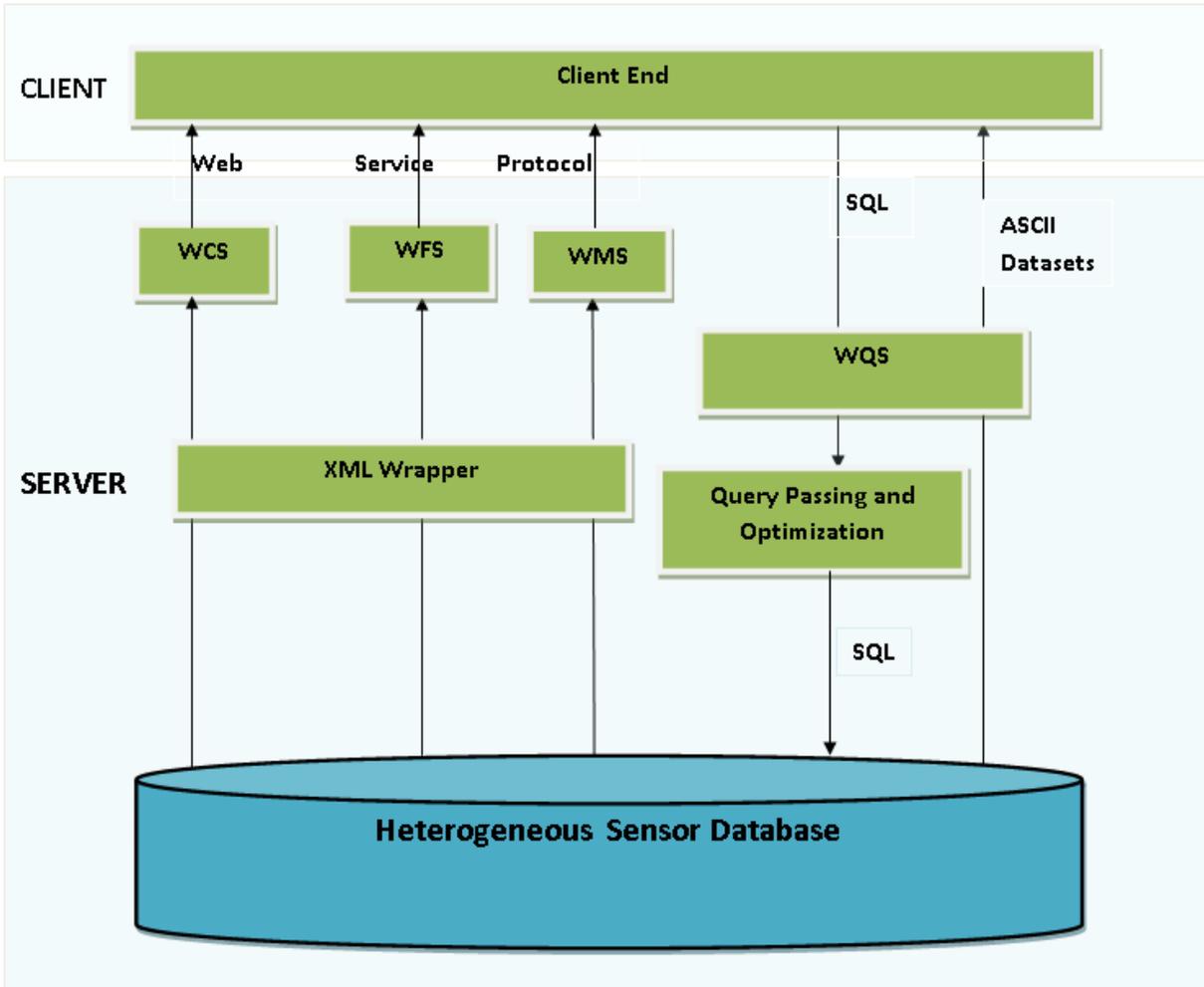

**Figure 6: Proposed Conceptual Architecture of Integrating the Heterogeneous Database and the Web Services**

## 5. Prototyping the Framework

In this section we carried out a prototypical implementation of the heterogeneous database model in PostGIS as displayed in figure 7. We loaded the tables with in-situ and remote sensor data as described in the logical model. In-situ sensor observations stored as vector coverages and remote sensor observations as raster coverages. On the database we had in-situ and remote sensor Land Surface Temperature LST coverage, Sea Surface Temperature SST coverage, Reference Evapotranspiration in-situ coverage, Normalised Difference Vegetation Index NDVI coverage and so on. Afterwards some few scenarios where the proposed heterogeneous sensor database model are leveraged to accomplish geo-scientific queries and processing involving remote and in-situ observations were tested. The query scenarios were executed from a client desktop application (the OpenJUMP Desktop GIS application) having established a connection to the heterogeneous sensor database server. Scenarios ranging from a simple case of obtaining the temperature difference between in-situ and remote temperature observations to a more complex case of estimating daily plant Evapotranspiration of a particular location were tested.

*5.1 Scenario 1: In-situ and satellite surface temperature analysis*

This scenario calculates the temperature difference between the in-situ sensor land surface temperature observation and remote sensor land surface temperature observation of a particular location. Listing 8 was used to obtain the required result from within the OpenJump desktop application.

Listing 1: Scenario 1 implementation SQL code

```sql
SELECT val1, (gv).val AS val2 ,val1-(gv).val AS diffval,geom

FROM ( SELECT ST_intersection(rast,the_geom) AS gv,

temp_value AS val1, ST_AsBinary(the_geom) AS geom

FROM in_situ_lst , lst_day

WHERE the_geom && rast

AND ST_intersects(rast,the_geom)

AND temp_lst_id = 1

) foo;
```

Here, this query picks up a particular temperature observation from the in-situ land surface temperature 'val1', in-situ_lst table of a location where id = 1, compares the temperature value with the corresponding remotely observed temperature, 'val2' of that same location on the raster temperature coverage, Lst_day and returns the difference, 'diffval'.

Figure 8 below is the implementation screen short excerpt from the OpenJump desktop application showing the connection to the heterogeneous sensor database and the result of the query from within the OpenJUMP client desktop application.

In figure 8 below, connection to the heterogeneous sensor database and the SQL query are depicted on the upper right hand side of the image while the query result on the lower left corner.

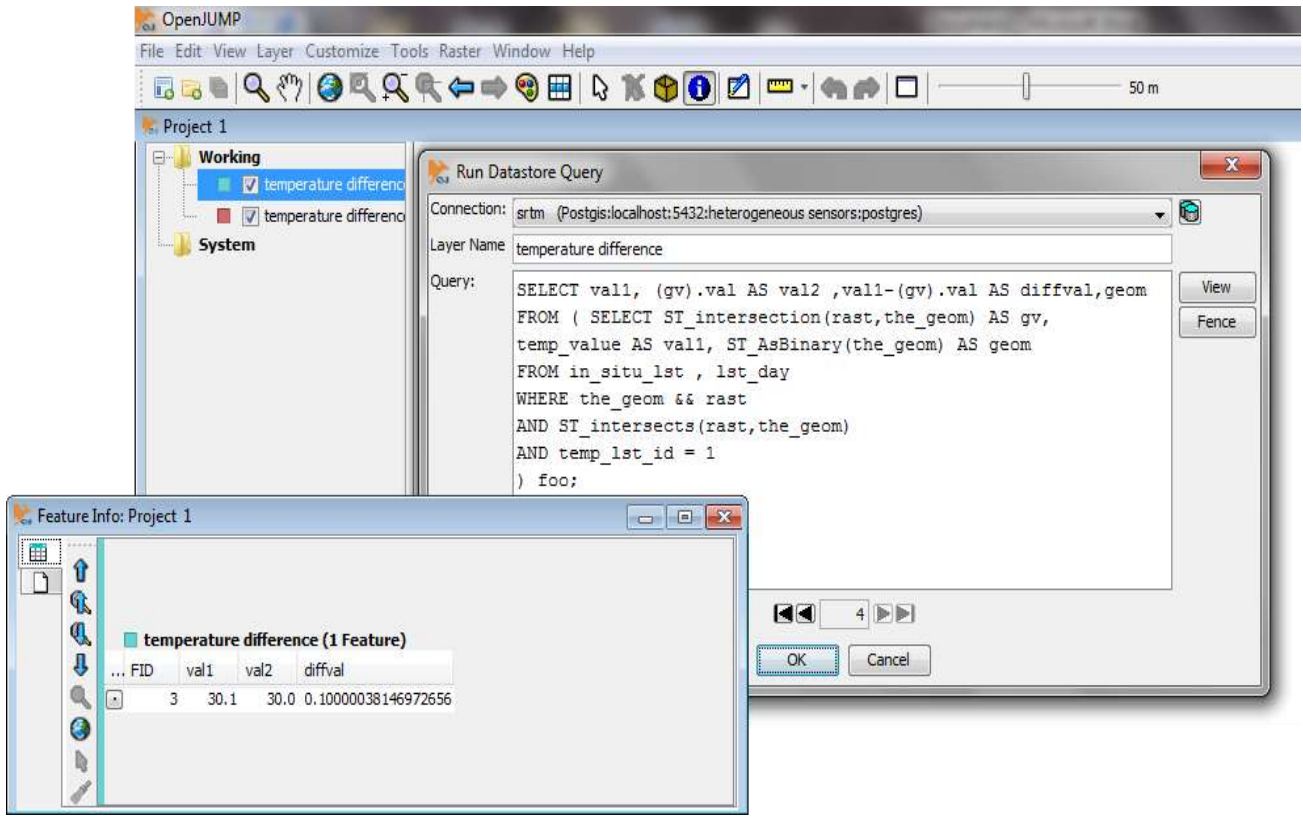

Figure8: Screen short of Scenario 1 implementation in OpenJump

## 5.2 Scenario2: Estimation of Actual Crop Evapotranspiration ET at the Database Backend

We have the in-situ Reference Evapotranspiration RET coverage from weather automatic stations and NDVI raster coverage of the spatio-temporal attribute in the heterogeneous sensor database. Therefore we can calculate the Actual Evapotranspiration AET, from an aggregation of RET and Fraction of Vegetation cover FVC, where FVC is derived from NDVI (Cutrell & Soylu, 2009).

To implement this scenario, we could use 1 and 0 as the approximate maximum and minimum NDVI values respectively within the area, this would give us an approximate estimation not very precise. But to obtain the actual NDVImax and NDVImin of the coverage area, we used the SQL query below in listings 2 and 3, which can then be factorized in the comprehensive AET query statement in listings 4 to obtain the precise AET.

Listing 2: SQL Query to obtain the NDVImax

```
SELECT (stats).max
FROM (SELECT ST_SummaryStats(rast) As stats
   FROM ndvi
   ORDER by stats DESC
   limit 1 ) As foo;
```

Listing 3: SQL Query to obtain the NDVImin

```
SELECT (stats).min
FROM (SELECT  ST_SummaryStats(rast) As stats
   FROM ndvi
   ORDER by stats ASC
   limit 1 ) As foo;
```

In this our example case, we calculated the AET of a point in the RET, in_situ_ret table where id =1 by implementing the SQL statement in listing 14 below. From the query in listing 21 and 22, we obtained the NDVImax and NDVImin as 0.86 and 0 respectively and factorized them in as shown in the listing 4 below and got the results from within the OpenJUMP client desktop application shown in figure 9.

Listing 4: Scenario 4 implementation SQL code

```
SELECT RET, NDVIp,(pow(((NDVIp-0.86)/(0.86-0)),2)) as FVC,
(pow(((NDVIp-0.86)/(0.86-0)),2))*RET as AET, the_geom
FROM (SELECT ST_Value(R.rast,I.the_geom) as NDVIp, I.value as RET,
ST_AsBinary(I.the_geom) as the_geom
FROM in_situ_ret I, ndvi R
WHERE ret_id = 1
AND  ST_Value(R.rast,I.the_geom) IS NOT NULL) foo;
```

Figure 9: Screen short excerpt of a sample Scenario 4 implementation result in OpenJump client end

Lots of other scenarios were also tested for example, calculation of Weighted Mean surface temperature values from a vector buffer. A scenario where, one could select a particular observation in the in-situ temperature observation, create a buffer of a given radius around that observation, then overlap this buffer geometry on the raster temperature coverage and obtains a weighted mean surface temperature value within the buffered region from the raster coverage. Also a scenario to describe a raster coverage metadata such as done in the OGC Sensor Web "DescribeCoverage" to obtain the metadata of a particular coverage. . In this case, we could leverage the PostGIS raster metadata description function to provide the client side description of a raster coverage through an SQL query.

In general the results of the sample queries shown above for the mentioned scenarios are alphanumeric or CSV formatted. They are returned to the client directly from the database. Other results formats are also possible as described in section 4.1 above, depending on how the client wants the results delivered.

*6. Evaluation and Conclusion*

*6.1 Query Flexibility*
The various geo-processing scenarios we implemented in the prototypical implementation exercise from within the OpenJump client side desktop application show that, this approach of delivering SQL based queries from the client end direct to the database backend makes it more flexible for the user on the client end to deliver geo-processing queries of extensive complexity involving in-situ and remote sensor observations. Language based query request such as the (SQL) has been considered advantageous especially by the database community because is very flexible, declarative, optimizable and more safe in evaluation (Baumann P. , 2009)This extensive support for different kinds of geo-processing and analysis involving in-situ and remote sensor observations through native SQL queries makes this approach advantageous to the current approach of having different geo-processing modules on the web service for specific purposes. In that case users are restricted only to the specific geo-processing capabilities the service offers.

### 6.2 Reduced communication load

The prototypical implementation our proposed heterogeneous sensor database model and the model of how it can be integrated seamlessly with the OGC geo-web services show that these disparate sensor observations can be integrated and managed in a single spatial database leveraging PostGIS 2.0 functionalities. Hence communication load to different databases are invariably reduced. The communication time lag incurred in the downloading of raster images from a flat file database via the ftp, obtaining in-situ observations from an in-situ sensor observation service and integrating the two on the service middleware level is invariably reduced greatly, adopting this heterogeneous database approach.

### 6.3 Work and Massive Data Retrieval Load

Also taking a look at the contents and the processes in dynamic systems such as in (Rueda & Gertz, 2008) (Kooistra, Bergsma, Chuma, & de Bruin, 2009), (AlastairAllen, Hamre, Pettersson, Murshed, & Sputh,2009), (Teillet, et al., 2007)and in the OGC Web Processing Services, they provide clients access and results based on pre-programmed calculations and/or computation models that operate on the spatial data. To enable geospatial processing and operations of diverse kinds, from simple subtraction and addition of sensor observations (e.g. the difference between satellite observed temperature and in-situ observation of a location) to complicated ones such as climate change models, requires the development of a large variety of models on the service middleware. This is massive in work load and huge amount of programming on the service. Also the data required for these services are usually retrieved dynamically from different databases and services which most times entails massive data retrieval especially from the satellite data (raster) storage.

Contrarily, by the means of a heterogeneous sensor database model such as developed and implemented in this research leveraging the functionalities of PostGIS 2.0 database extension, geo-processing and analytics involving remote and in-situ sensor data are carried out at the database backend by native SQL request statements. Therefore the variety of geo-processing work load on the service middleware is reduced. The service middleware in our case is majorly for service delivery from the client to database and vice versa. Massive data retrieval before processing is completely avoided. Also massive programming involved in the development of different kinds of geo-processing models on the web service is reduced.

## 7. *Further Work*

The practical usefulness of this proposed approach will be very more appreciated and leveraged when we are done with the full implementation of the model, integrating the proposed sensor heterogeneous database and other geo-web services in the SOS. This is our next milestone, to fully integrate this heterogeneous sensor database framework and the WQS with other geo-web services for query result delivery to the clients in different formats as described in figure 6.